\documentclass[%
 aip,
 amsmath,amssymb,
 reprint,%
cha
]{revtex4-1}

\usepackage{graphicx}
\usepackage{dcolumn}
\usepackage{bm}

\usepackage[utf8]{inputenc}
\usepackage[T1]{fontenc}
\usepackage{mathptmx}
\usepackage{etoolbox}

\makeatletter
\def\@email#1#2{%
 \endgroup
 \patchcmd{\titleblock@produce}
  {\frontmatter@RRAPformat}
  {\frontmatter@RRAPformat{\produce@RRAP{*#1\href{mailto:#2}{#2}}}\frontmatter@RRAPformat}
  {}{}
}%

\usepackage{xcolor}

\definecolor{deep_blue}{RGB}{31, 119, 180}
\definecolor{deep_orange}{RGB}{255, 127, 14}
\definecolor{deep_green}{RGB}{44, 160, 44}
\definecolor{deep_red}{RGB}{214, 39, 40}
\definecolor{deep_purple}{RGB}{148, 103, 189}
\definecolor{deep_brown}{RGB}{140, 86, 75}
\definecolor{deep_pink}{RGB}{227, 119, 194}

\usepackage[colorlinks,linktoc=all]{hyperref}
\hypersetup{
  linkcolor=deep_pink,    
  citecolor=deep_green,   
  urlcolor=deep_red       
}

\makeatother
\begin{document}

\preprint{AIP/123-QED}

\title[Minimal Deterministic Echo State Networks Outperform Random Reservoirs in Learning Chaotic Dynamics]{Minimal Deterministic Echo State Networks Outperform Random Reservoirs in Learning Chaotic Dynamics}
\author{F. Martinuzzi}
    \affiliation{Center for Scalable Data Analytics and Artificial Intelligence (ScaDS.AI), Leipzig University, Leipzig, Germany}
    \affiliation{Institute for Earth System Science \& Remote Sensing, Leipzig University, Leipzig, Germany}
    \email{martinuzzi@informatik.uni-leipzig.de}

\date{\today}

\begin{abstract}

    Machine learning (ML) is widely used to model chaotic systems. Among ML approaches, echo state networks (ESNs) have received considerable attention due to their simple construction and fast training. However, ESN performance is highly sensitive to hyperparameter choices and to its random initialization. In this work, we demonstrate that ESNs constructed using deterministic rules and simple topologies (MESNs) outperform standard ESNs in the task of chaotic attractor reconstruction. We use a dataset of more than 90 chaotic systems to benchmark 10 different minimal deterministic reservoir initializations. We find that MESNs obtain up to a $41\%$ reduction in error compared to standard ESNs. Furthermore, we show that the MESNs are more robust, exhibiting less inter-run variation, and have the ability to reuse hyperparameters across different systems. Our results illustrate how structured simplicity in ESN design can outperform stochastic complexity in learning chaotic dynamics.
    
\end{abstract}

\maketitle

\begin{quotation}

    Echo state networks (ESNs) offer an appealing machine learning formula for modeling chaotic systems: a fixed, high-dimensional reservoir of interconnected units remains fixed, a read-out layer is trained through linear regression, and the final network is able to replicate the system’s long-term behaviour. The issue is that the reservoir is usually randomly generated. As a result, two ESNs built with the same hyperparameters can perform very differently. Additionally, finding a good configuration often feels more like trial and error than design. In this work, we ask whether randomness is really necessary for reconstructing chaotic dynamics with ESNs. We benchmark minimal deterministic reservoirs, whose connections follow simple fixed rules instead of randomness, on multiple different chaotic systems. We find that these minimal deterministic reservoirs consistently outperform random initializations. By removing randomness and reducing the burden of hyperparameter tuning, minimal deterministic ESNs point toward more reliable models for chaotic time series reconstruction.
    
\end{quotation}

\section{\label{mesncd:sec:intro} Introduction}
    Chaos is prevalent in both natural and engineered systems \citep{ott2002chaos, strogatz2018nonlinear}. In many of these circumstances, access to the underlying equations generating the data is not possible. Therefore, to study the dynamical system of interest, it is necessary to build a surrogate of the model from time series alone. However, the complex temporal structures of chaotic dynamics, known as attractors, are notoriously difficult to reconstruct \citep{packard1980geometry, takens1981detecting, abarbanel1993analysis, kantz2003nonlinear}. Unlike forecasting, which aims to predict short-term behavior, attractor reconstruction focuses on building generative models that reproduce the long-term statistical properties of a system, i.e., its \emph{climate} \citep{patel2023using, platt2023constraining, hemmer2025true}. The data-driven nature of this problem has led recent approaches to increasingly rely on machine learning (ML) methods to deal with chaotic data. \citep{brunton2022datadriven, durstewitz2023reconstructing, gilpin2024generative}.
    
    Echo state networks (ESNs) have been successfully used to reconstruct chaotic attractors \citep{pathak2017using, lu2018attractor, fan2020longterm, rhm2021modelfree, zhang2021learning, scully2021measuring, gauthier2022learning, smith2022learning, margazoglou2023stability}. Generally belonging to the family of reservoir computing (RC) models \citep{verstraeten2005reservoir, verstraeten2007experimental}, ESNs are a type of recurrent neural network (RNN) that is not trained through backpropagation \citep{jaeger2002tutorial}. Instead, ESNs expand input data into a higher-dimensional space via a \emph{reservoir}, which typically consists of a randomly connected RNN with fixed weights \citep{jaeger2001echo, jaeger2004harnessing}. The resulting high-dimensional representations, called \emph{states}, are then used to train only the output layer, usually through linear regression \citep{lukoeviius2012practical}. Owing to their simple architecture, ESNs have gained popularity in the nonlinear dynamics community as an accessible ML tool for dynamical system reconstruction \citep{bollt2021explaining, panahi2024adaptable}. Their increased adoption is reinforced by a growing theoretical literature showing that ESNs possess strong inductive biases for learning chaotic systems \citep{grigoryeva2018echo, hart2020embedding, hart2021echo, grigoryeva2021chaos, grigoryeva2023learning}.
    
    However, the performance of ESNs is highly sensitive to hyperparameter tuning and dependent on a correct initialization of the random weights. In fact, even changing a single hyperparameter results in a failed reconstruction of chaotic attractors \citep{pathak2017using, chen2023proper}. Even with optimal hyperparameters, different random seeds can result in very different reservoir matrices, which lead to reconstructions of varying quality \citep{haluszczynski2019good}. While considerable work has been done to design alternative approaches to ESN construction and tuning \citep{livi2018determination, griffith2019forecasting, racca2021robust, grigoryeva2024datadriven, hart2024attractor}, the process still relies heavily on heuristics. Additionally, most proposed solutions do not remove the inherent randomness involved in creating ESNs. Minimal complexity ESNs (MESNs  \citep{rodan2011minimum}) provide an alternative approach to building ESNs, which eliminates random elements from the process. Unlike standard ESN initializations, these minimalist reservoirs follow simple deterministic rules and assign identical-magnitude weights of the same sign to all internal connections.
    
    Do minimally deterministic ESNs outperform randomly initialized reservoirs in reconstructing chaotic attractors? Although MESNs' topologies are supported by emerging theoretical insights \citep{li2024simple, fong2025universality}, they have received far less practical attention compared to conventional ESNs. Nevertheless, studies indicate that small reservoirs with minimal or deterministic connectivity can model chaotic dynamics \citep{carroll2019network, griffith2019forecasting, ma2023efficient, ma2023novel, jaurigue2024chaotic, viehweg2025deterministic}. In this work, we systematically compare several MESN constructions with standard random ESNs on more than ninety distinct chaotic systems. Our study evaluates whether minimal deterministic initializations can provide a reliable, reproducible, and potentially better approach for modeling chaotic dynamics.
    
\section{\label{mesncd:sec:meth} Task and Methods}

    To test the ability of MESNs and the ESN to learn chaotic dynamics, we train them to perform autoregressive forecasting. Formally, let $\{\mathbf{y}_t\}_{t=1}^T$ be a multivariate timeseries from a chaotic system with $\mathbf{y}_t \in \mathbb{R}^D$. The goal is to generate a time series with the same properties as $\mathbf{y}_{T+1}, \dots, \mathbf{y}_{T+H}$ by learning a mapping $\mathcal{M}: \mathbb{R}^{D} \rightarrow \mathbb{R}^D$ such that $\hat{\mathbf{y}}_{T+1} = \mathcal{M}(\mathbf{y}_T)$, $\hat{\mathbf{y}}_{T+2} = \mathcal{M}(\hat{\mathbf{y}}_{T+1})$, and so on recursively up to $\hat{\mathbf{y}}_{T+H} = \mathcal{M}(\hat{\mathbf{y}}_{T+H-1})$, where $H$ is the forecast length. The performance of $\mathcal{M}$ depends on certain configuration values that govern key aspects of the model, known as \emph{hyperparameters}. Finding the right hyperparameters is essential for building an effective model. In this work, $\mathcal{M}$ represents either an ESN or MESNs. We fix $D = 3$, as all systems considered are three-dimensional.

    We use Julia \citep{bezanson2017julia} to run all the simulations, and we generate the figures using \texttt{Makie.jl} \citep{danisch2021makiejl}. We use a Dell XPS model 9510 laptop with an Intel Core i7-11800H CPU and 16 GB of RAM for all experiments. Given the low dimensionality of simulations, we use no GPU acceleration.

    \subsection{\label{mesncd:subsec:esn} Echo State Networks}
        ESNs are an ML model proposed by \citet{jaeger2004harnessing} where the internal weights are generated randomly and kept fixed. The update equations of the ESN are as follows:
        \begin{equation}
        \label{mesncd:eq:esn}
            \mathbf{x}(t) = \tanh\left(\mathbf{W}_\text{in} \mathbf{u}(t) + \mathbf{W} \mathbf{x}(t-1) + \mathbf{b} \right),
        \end{equation}
        where $\mathbf{x}(t) \in \mathbb{R}^N$ is the reservoir state at time $t$, and $\mathbf{u}(t) \in \mathbb{R}^D$ is the input signal. The matrices $\mathbf{W}_\text{in} \in \mathbb{R}^{N \times D}$ and $\mathbf{W} \in \mathbb{R}^{N \times N}$ define the input and reservoir weights, respectively, and $\mathbf{b} \in \mathbb{R}^N$ is a bias term. In all experiments, we set $D = 3$ and fix the reservoir size to $N = 300$.

        Since the weights $\mathbf{W}_\text{in}, \mathbf{W}, \mathbf{b}$ are not updated during training, their initialization is critical to performance. The entries of $\mathbf{W}_\text{in}$ are sampled from a uniform distribution $\mathcal{U}(-\sigma, \sigma)$, with $\sigma=0.1$ in this work. Following the architecture proposed by \citet{lu2017reservoir}, we construct $\mathbf{W}_\text{in}$ such that each of the input signals $M$ connects to the reservoir nodes $N/M$. The reservoir matrix $\mathbf{W}$ is typically sparse, with sparsity treated as a hyperparameter. The non-zero entries are drawn from $\mathcal{U}(-1, 1)$ and rescaled to have a spectral radius $\rho$. Tuning $\rho$ is considered crucial for predictive accuracy \citep{jiang2019modelfree}. The pattern of non-zero entries in $\mathbf{W}$ defines its topology. In this work, we use the terms "initialization" and "topology" interchangeably, as different initializations result in structurally distinct reservoirs. The bias vector $\mathbf{b}$ is initialized as zero in this work.

        During the training phase, the data is passed through the ESN, and the resulting reservoir states ${\mathbf{x}_1, \ldots, \mathbf{x}_T}$ are collected into a state matrix $\mathbf{X} \in \mathbb{R}^{T \times N}$.

        The prediction for step $t$ is given by
        \begin{equation}
        \label{mesncd:eq:esn_output}
            \hat{\mathbf{y}}(t) = \mathbf{W}_{\text{out}} \, \mathbf{x}(t),
        \end{equation}
        where $\mathbf{W}_{\text{out}} \in \mathbb{R}^{L \times N}$ is the output matrix and $L$ is the dimension of the output data. In our work, $L=D=3$. The output weights are computed using ridge regression (Tikhonov regularization) as:
        \begin{equation}
        \label{mesncd:eq:tikhonov}
            \mathbf{W}_{\text{out}} = \mathbf{Y} \mathbf{X}^\top \left( \mathbf{X} \mathbf{X}^\top + \beta \mathbf{I} \right)^{-1},
        \end{equation}
        where $\mathbf{Y} \in \mathbb{R}^{L \times T}$ contains the target outputs and $\beta$ is the regularization parameter.

        \begin{table}[ht]
        \centering
        \begin{tabular}{@{}lcc@{}}
            \toprule
            Hyperparameters & \textbf{ESN} & \textbf{Minimal ESNs} \\
            \hline
            Spectral Radius ($\rho$) & 0.7:0.1:1.3 & / \\
            Sparsity            & 0.01, 0.03, ..., 0.1 & / \\
            Reservoir Size      & 300 & 300 \\
            Weights             & $\mathcal{U}(-1.0, 1.0)$ & 0.1 \\
            Input Weights      & $\mathcal{U}(-0.1, 0.1)$ & $\pm0.01$ \\
            Regularization ($\beta$)      & 0, $10^{-1}$:$10^{-1}$:$10^{-10}$ & $10^{-1}$:$10^{-1}$:$10^{-17}$ \\
            \hline
        \end{tabular}
        \caption{\textbf{Comparison of hyperparameter configurations for ESNs and MESNs.} The notation for parameters range follows the convention start:step:stop. The ESNs vary several key parameters, including spectral radius, sparsity, and randomly initialized reservoir weights. MESNs use fixed, deterministic weights (set to 0.1) and do not require tuning of spectral radius or sparsity. Both models use reservoirs of size 300. Regularization coefficients ($\beta$) are explored over a broader range for MESNs. Input weights are either sampled from a uniform distribution (ESN) or fixed with changing signs (MESN).}
        \label{mesncd:table:hp}
        \end{table}

    \subsection{\label{mesncd:subsec:mcesn} Minimum Complexity Echo State Networks}

        To reduce the complexity typically associated with tuning ESNs, \citet{rodan2011minimum} proposed a minimal approach to constructing the input and reservoir matrices. This method, known as minimum complexity ESNs (MESNs), eliminates the inherent randomness in the initialization of ESN matrices. Subsequent work extended the range of viable reservoir topologies within this framework \citep{rodan2012simple, elsarraj2019demystifying, fu2023doublecycle}. Despite their differences, all proposed input and reservoir configurations follow the same core principles.
        
        In MESNs, all the entries in the input matrix $\mathbf{W}_{\text{in}}$ have the same absolute value $a > 0$, differing only in sign. The sign can be assigned either randomly or deterministically, for example by sampling the decimal digits of an irrational number: if a digit is odd, the corresponding sign is negative; otherwise, it is positive. In this work, we fix $a=0.01$ and the sign of each weight is determined from sampling the decimals of $\pi$, proceeding in a columnwise fashion.

        On the other hand, the reservoir topologies use only weights with the same sign. We consider ten distinct topologies:
        \begin{enumerate}
            \item \textbf{Delay line (DL)}\citep{rodan2011minimum}: weights arranged in a line with feedforward weights $W_{i+1,i} = r$ on the subdiagonal.
            
            \item \textbf{Delay line with feedback (DLB)}\citep{rodan2011minimum}: adds feedback weights $W_{i,i+1} = b$ on the superdiagonal to the DL topology.
            
            \item \textbf{Simple cycle (SC)}\citep{rodan2011minimum}: weights form a cycle with weights $W_{i+1,i} = r$ and wrap-around connection $W_{1,N} = r$.

            \item \textbf{Cycle with jumps (CJ)}\citep{rodan2012simple}: Like SC, but with additional bidirectional jump connections of fixed distance $\delta$, all with weight $r_j$. Cycle weights are $W_{i+1,i} = r$ and $W_{1,N} = r$.

            \item \textbf{Self-loop cycle (SLC)}\citep{elsarraj2019demystifying}: A cycle reservoir with added self-loops. Nonzero weights are $W_{i+1,i} = r$, $W_{1,N} = r$, and $W_{i,i} = ll$.
    
            \item \textbf{Self-loop feedback cycle (SLFB)}\citep{elsarraj2019demystifying}: Cycle reservoir with $W_{i+1,i} = r$ and $W_{1,N} = r$; even-indexed units have feedback $W_{i,i+1} = r$, odd ones have self-loops $W_{i,i} = ll$.
            
            \item \textbf{Self-loop delay line with backward connections (SLDB)}\citep{elsarraj2019demystifying}: Forward path $W_{i+1,i} = r$, backward links $W_{i+2,i} = r$, and self-loops $W_{i,i} = ll$ for all units.
            
            \item \textbf{Self loop with forward skip connections (SLFC)}\citep{elsarraj2019demystifying}: Only skip connections $W_{i+2,i} = r$ (no standard forward links), plus self-loops $W_{i,i} = ll$.
            
            \item \textbf{Forward skip connections (FC)}\citep{elsarraj2019demystifying}: Only $W_{i+2,i} = r$ are non-zero.

            \item \textbf{Double cycle (DC)}\citep{fu2023doublecycle}: Like SC, but with connections on both subdiagonal and superdiagonal. That is, $W_{i+1,i} = r$ and $W_{i,i+1} = r$, plus the wrap-around connections $W_{1,N} = r$ and $W_{N,1} = r$.

        \end{enumerate}
        In this work we further simplify the setup by setting all relevant weights to the same constant value: $r = b = ll = r_j = 0.1$. Furthermore, the bias $\mathbf{b}$ in Eq.~\ref{mesncd:eq:esn} is initialized as a vector of ones for MESNs.

        We use the software \texttt{ReservoirComputing.jl} \citep{martinuzzi2022rcjl} for the implementation and training of the ESNs and MESNs.
        
    \subsection{\label{mesncd:subsec:metrics} Accuracy Metric - Fractal Dimension}

        To determine whether a model has learned the chaotic dynamics of a given system, we want to see if it can replicate its attractor. The long-term evolution of a chaotic attractor in the phase space can be characterized by its fractal dimension \citep{falconer2013fractal}. Following prior studies \citep{zhang2025zeroshot, lai2025panda}, we rely on the method proposed by \citet{grassberger1983characterization}, which estimates the correlation dimension. To assess the correct reconstruction of a chaotic attractor, we compute the absolute difference between the real correlation dimension and the correlation dimension of the forecasted system (CDE). We use the Julia package \texttt{FractalDimensions.jl} \citep{datseris2023fdjl} to compute the correlation dimension.

    \subsection{\label{mesncd:subsec:data} Chaotic Systems Dataset and Hyperparameter Tuning}

        We source the chaotic systems from the chaotic system dataset \texttt{dysts} \citep{gilpin2021chaos}, accessed through the Julia port \footnote{https://github.com/nathanaelbosch/ChaoticDynamicalSystemLibrary.jl}. To uniform experiments, we exclude systems with more than three dimensions and those exhibiting integration instabilities, resulting in a final set of 91 systems. All systems are integrated using the \texttt{Feagin12} solver from the \texttt{DifferentialEquations.jl} package \citep{rackauckas2017dejl}, with absolute and relative tolerances set to $10^{-13}$. Following \citet{lu2017reservoir}, data is standardized to zero mean and unit variance during training and prediction, and rescaled back before computing accuracy metrics. For each system, we generate 20 time series from different initial conditions.

        We simulate 7,000 data points for training and 2,500 for testing for each system, discarding the first 300 points to eliminate transient dynamics. Model hyperparameters are tuned using temporal cross-validation \citep{bergmeir2012crossvalidation, bergmeir2018note}: the training data is split into \emph{folds} of increasing length, which preserve temporal order. Each fold is further divided into training and validation subsets. The model is trained on the training portion and evaluated on the corresponding validation set. After evaluating all folds, we select the hyperparameters that achieve the best overall validation performance. In this work, we use four folds with a fixed validation size of 350 time steps. Optimal hyperparameters are selected via grid search over the space defined in Table~\ref{mesncd:table:hp}. Once selected, each model is retrained on the full 7,000-point training set and used to generate 2,500 test points in an autoregressive manner.

\section{\label{mesncd:sec:res} Results}

    \subsection{\label{mesncd:subsec:mp} Deterministic Reservoirs Outperform Random Reservoirs}

        \begin{figure}[ht]
            \centering
            \includegraphics[width=\columnwidth]{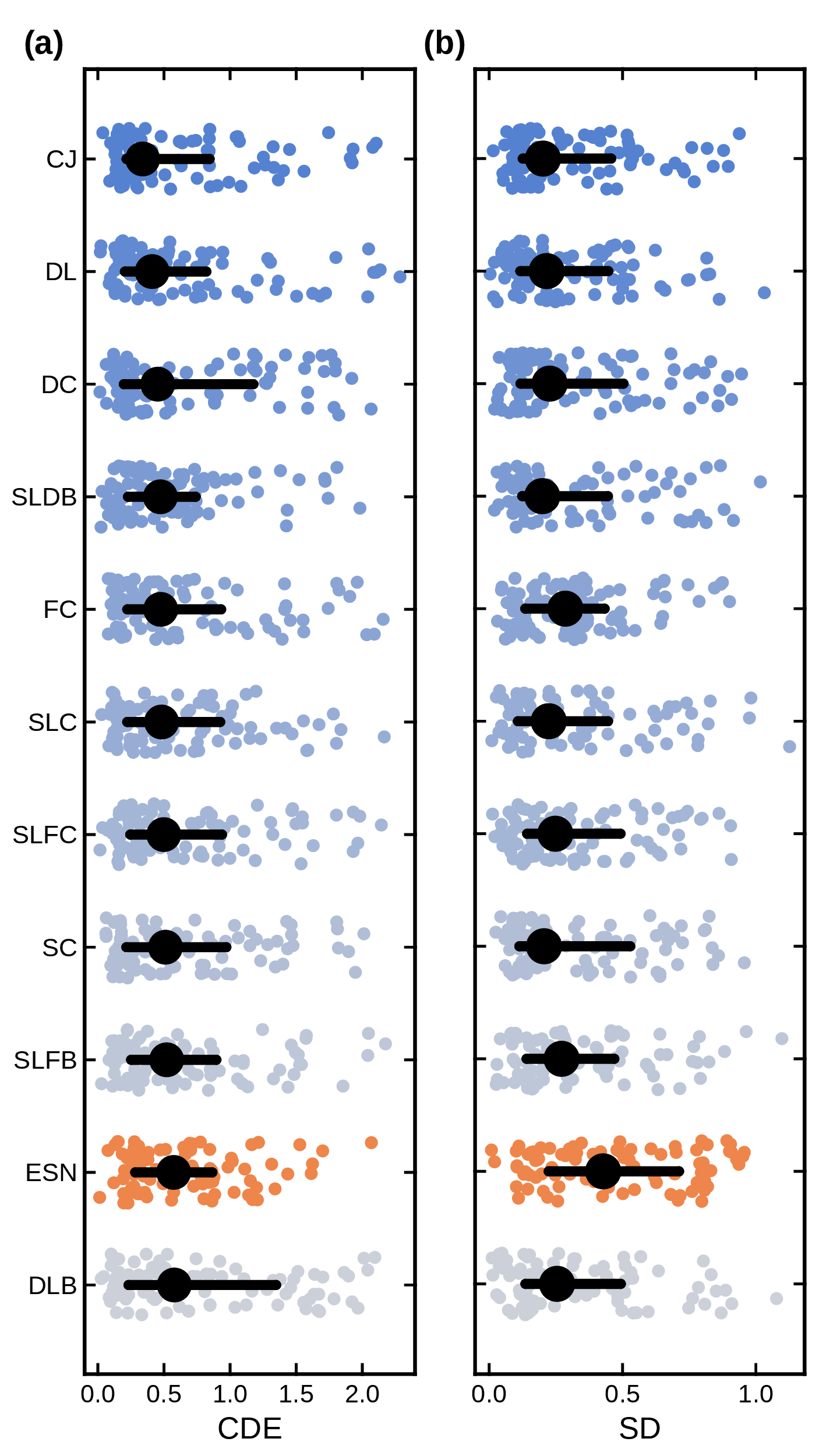}
            \caption{\textbf{Performance comparison of minimal deterministic reservoir topologies and standard ESNs in chaotic attractor reconstruction.} Panel (a) shows the average absolute error between the true and predicted correlation dimension (CDE) for each system over 20 realizations, shown for 10 minimal topologies and the standard ESN. Panel (b) shows the corresponding standard deviation (SD) of the CDE across the same realizations. Each dot represents the average result for a single system; black markers indicate the median value for each initializer, with error bars showing the interquartile range. 
            }
            \label{mesncd:fig:cde}
        \end{figure}

        Figure~\ref{mesncd:fig:cde}a compares MESNs initialized with 10 different minimal deterministic reservoir topologies against a standard ESN with classic random initialization. Each point represents the average CDE for system, over all systems. Black markers indicate the median error for each topology, with accompanying error bars.
        
        The CJ topology achieves the lowest median error of 0.34, consistent with its strong theoretical foundations \citep{rodan2012simple, li2024simple}. The simplest topology in this study, DL, ranks second with a median CDE of 0.41. In contrast, the ESN with random initialization shows the second-worst performance, with a median error of 0.57. We find that CJ achieves a CDE below 0.5 on 57\% of systems, compared to 48\% for the ESN. While most CJ errors are concentrated below 0.5, the ESN exhibits a broader error spread, ranging from 0.0 to 1.5. The ESN shows fewer extreme outliers, with only one case exceeding an error of 2.0, whereas deterministic reservoirs tend to produce more outliers. However, we note that beyond a CDE of approximately 1.6, reconstructions fail catastrophically, resulting in distinctions among larger errors being qualitatively irrelevant.

        The differences between MESNs and ESNs become even more pronounced when we examine the standard deviation (SD) of the CDE. Figure~\ref{mesncd:fig:cde}b shows the SD over 20 different initial conditions for each system and reservoir initialization. Deterministic reservoirs exhibit relatively low and consistent variability, with SD values ranging from 0.12 (SLDB) to 0.25 (DLB); SLFB is a slight outlier at 0.27. In contrast, the ESN shows a higher median SD of 0.43. Although deterministic topologies still produce more outliers, the contrast is less pronounced than in Figure~\ref{mesncd:fig:cde}a. Additionally, the ESN’s SD distribution is more dispersed than its error distribution in panel (a), indicating that the randomness in its initialization contributes to greater variability in performance across runs.

    \subsection{\label{mesncd:subsec:cr} Deterministic Reservoirs are Robust and Reusable}
    
        \begin{figure*}[ht]
            \centering
            \includegraphics[width=\textwidth]{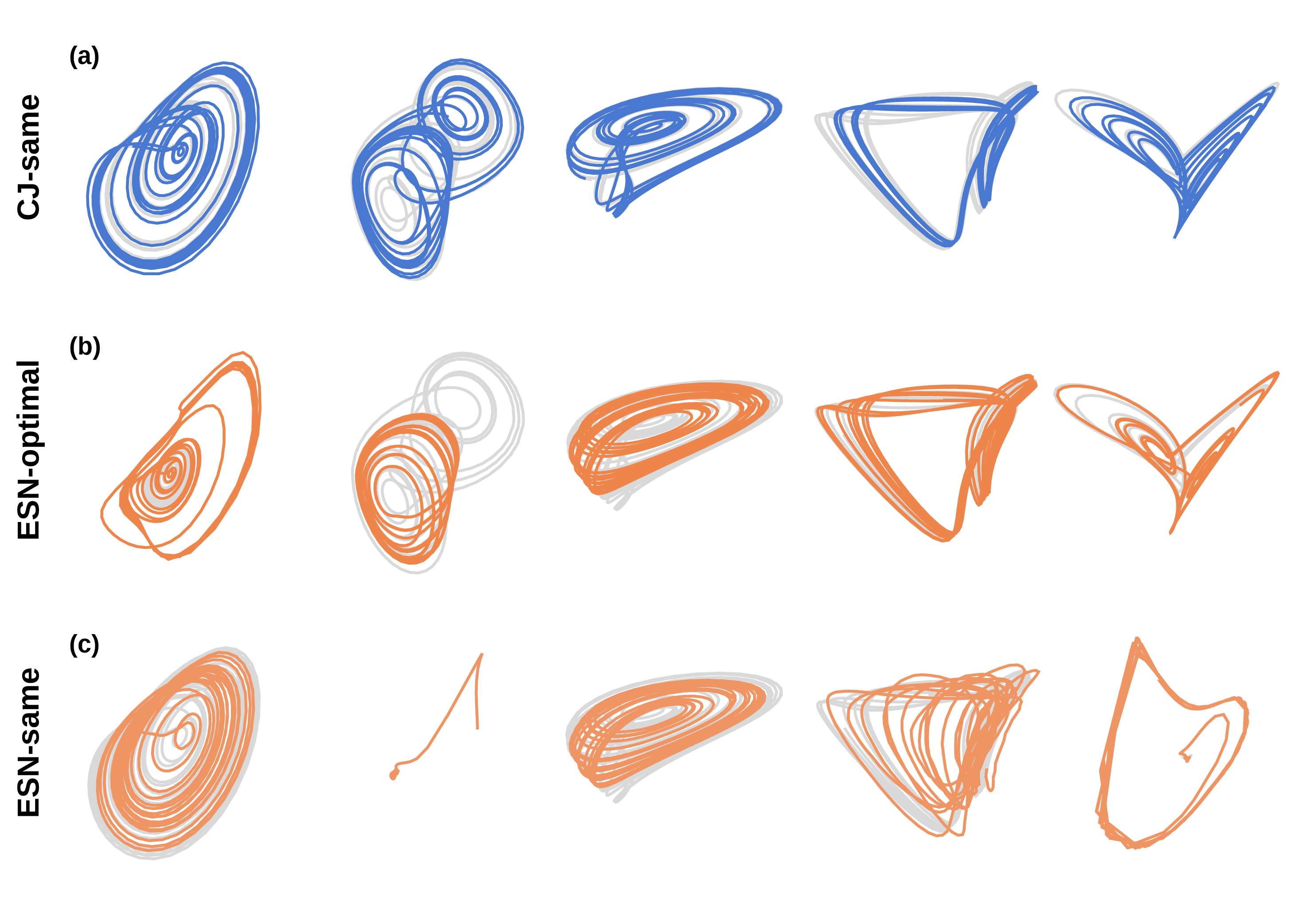}
            \caption{\textbf{Comparison of attractor reconstructions for five chaotic systems using different reservoir initializations and hyperparameter settings.} In this figure, each column corresponds to a different system; gray traces represent the true attractors, and colored traces represent the reconstructions. Each row corresponds to a unique initializer with an associated hyperparameter strategy. Panel (a) shows MESNs initialized with the CJ topology, using identical hyperparameters on all five systems. In panel (b), we show ESNs with random reservoir initialization, using individually optimized hyperparameters for each system. Finally, in panel (c), we show ESNs using the same hyperparameters for all systems, chosen to be optimal for the third system.}
            \label{mesncd:fig:comp}
        \end{figure*}

        In Figure~\ref{mesncd:fig:comp} we compare the reconstruction performance of MESNs initialized with the CJ topology and standard ESNs for five different chaotic systems. Each column corresponds to a different system. For clarity, only the first 900 steps of the 2500-step prediction are shown in each case. All reported error values refer to this 900-step segment unless otherwise noted.

        The first row, Figure~\ref{mesncd:fig:comp}a, shows reconstructions using the CJ topology with identical (optimal) hyperparameters. CJ demonstrates consistently good performance on all five systems. The resulting CDEs, from left to right, are 0.16, 0.06, 0.28, 0.19, and 0.30. When the forecast length is extended to 2,500 steps, these values change only modestly. None of the changes affect the overall quality of the reconstruction, suggesting that CJ returns stable and reliable long-term forecasts under fixed hyperparameter settings.

        In the middle row, Figure~\ref{mesncd:fig:comp}b shows reconstructions of the same five chaotic systems using ESNs with individually optimized hyperparameters. As with CJ, the resulting CDEs suggest generally good reconstructions: 0.20, 0.03, 0.33, 0.08, and 0.13 (from left to right). However, a closer visual inspection reveals limitations not captured by these error values. For the second system, the ESN fails to reconstruct the full structure of the attractor, unlike the CJ model, despite achieving a very low error. Additionally, for the fourth system, the ESN suffers a catastrophic failure when the prediction length is extended to 2,500 steps: the error increases from 0.08 to 1.4. This rise indicates that, at some point during the forecast, the model prediction deviates from the true attractor.

        Finally, the third row, Figure~\ref{mesncd:fig:comp}c, shows the attractor reconstructions produced by ESNs, all using the same set of hyperparameters optimized for the third system. The results show several reconstruction failures, with CDEs of 0.48, 1.72, 0.08, 1.70, and 1.60, from left to right. These outcomes illustrate the point discussed earlier: CDEs $>1.6$ consistently correspond to catastrophic reconstruction failures. This is clearly observed for the second and fifth systems, where the predicted dynamics diverge substantially from the true attractors. 

        These results indicate that for MESNs, it is possible not only to use the same reservoir topology to model different chaotic systems, but also to keep the same hyperparameters. In fact, for the reconstructions shown in Figure~\ref{mesncd:fig:comp}a, the only difference between the models lies in the output layer. All other components, including the input and reservoir matrices, weights, and hyperparameters, are identical. It is important to note that in Figure~\ref{mesncd:fig:comp} we only display one model run per system. In the case of CJ, this does not matter, since all realizations are identical. However, in the case of the ESN, different runs can have a large impact on the reconstruction abilities, as discussed in Section~\ref{mesncd:subsec:mp}.
        

    \subsection{\label{mesncd:subsec:hp} Sensitivity to Hyperparameters of Deterministic Reservoirs}

        \begin{figure}[ht]
            \centering
            \includegraphics[width=\columnwidth]{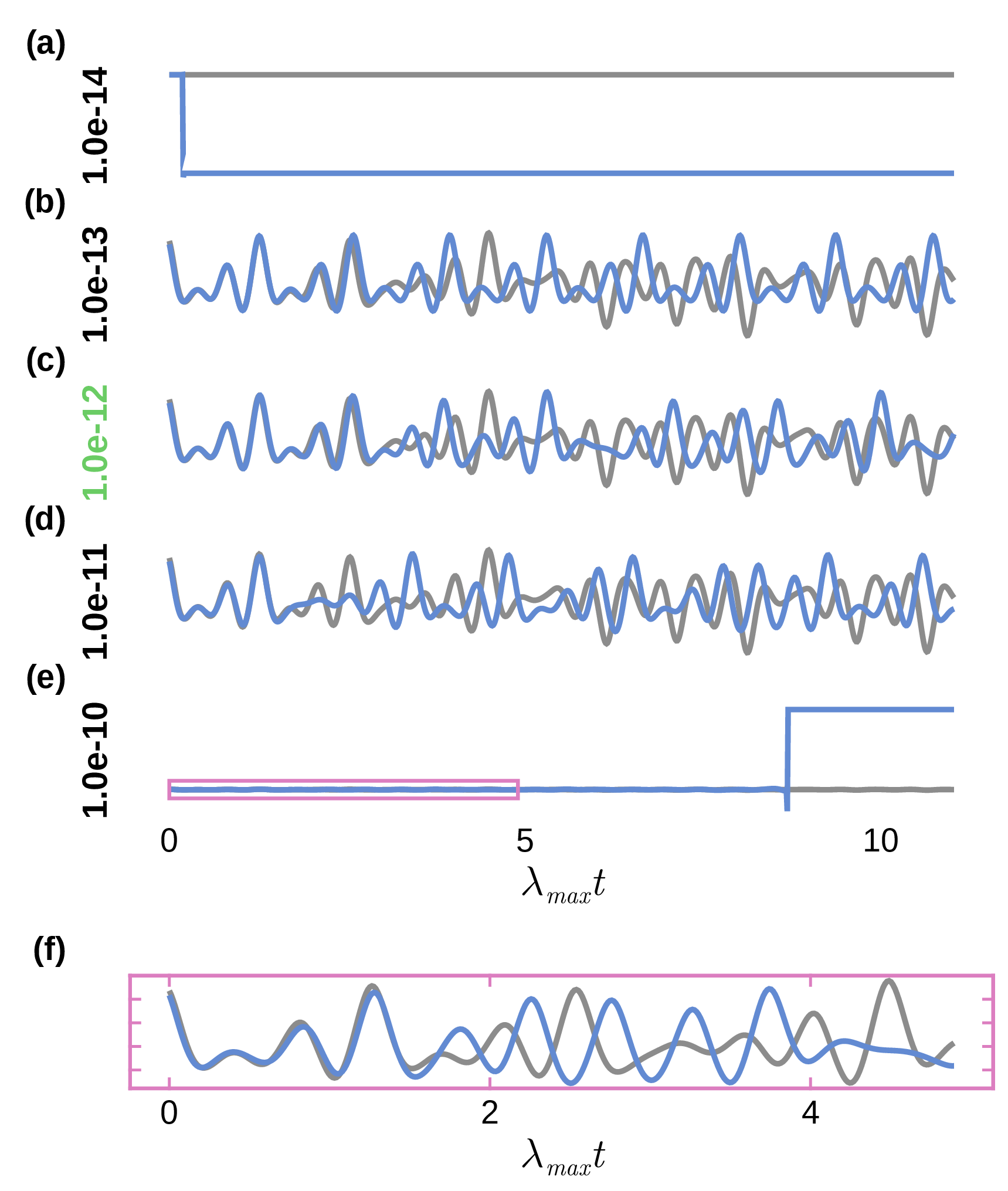}
            \caption{\textbf{Effect of regularization parameter $\beta$ on the reconstruction quality of a chaotic time series using a MESN initialized with the DL topology.} Each panel shows the true $y$-component of a chosen chaotic system (gray) and the predicted output (blue) over 900 normalized time steps showcased in Lyapunov time ($\lambda_{\max} t$). Panels (a)–(e) correspond to decreasing levels of regularization, from $\beta = 1.0 \times 10^{-14}$ to $1.0 \times 10^{-10}$, with the optimal value ($\beta = 1.0 \times 10^{-12}$) highlighted in green in (c). Panel (f) provides a zoomed-in view of the early portion of the forecast in (e), illustrating how short-term predictions can appear accurate even when long-term reconstruction fails. Abrupt divergence in panels (d) and (e) illustrates how overly large regularization values lead to catastrophic failure in attractor reconstruction despite initially low error.}
            \label{mesncd:fig:ts}
        \end{figure}

        For MESNs, the only tuned hyperparameter is the regularization coefficient $\beta$. Having previously analyzed the impact of hyperparameters on ESN performance (Section~\ref{mesncd:subsec:cr}), we now examine how varying $\beta$ affects MESN reconstruction. Figure~\ref{mesncd:fig:ts} shows the $y$-component of a chaotic system alongside reconstructions from a DL-initialized MESN using five different $\beta$ values. The optimal value, highlighted in green, is $\beta = 1.0 \times 10^{-12}$; the others are scaled up and down by two orders of magnitude. For clarity, only the first 900 prediction steps are shown. Unless otherwise noted, all metrics are computed over this segment.

        In Figure~\ref{mesncd:fig:ts}a, we show the prediction obtained using a regularization parameter of $\beta = 1.0 \times 10^{-14}$. The MESN's failure to reproduce the dynamics of the system is clear, with failure occurring within the first 50 steps. The resulting CDE is 1.78.

        Closer to the optimal regularization value, Figure~\ref{mesncd:fig:ts}b shows the time series reconstructed by the DL-initialized MESN with $\beta = 1.0 \times 10^{-13}$. Unlike the previous case, the reconstruction is more accurate, with a reported CDE of 0.35.

        Figure~\ref{mesncd:fig:ts}c presents the results obtained using the optimal regularization value of $\beta = 1.0 \times 10^{-12}$. In this case, the CDE is 0.14, indicating an accurate reconstruction.

        With a regularization value of $\beta = 1.0 \times 10^{-11}$, Figure~\ref{mesncd:fig:ts}d shows the lowest CDE: 0.09. This occurs despite $\beta$ not being the optimal hyperparameter. However, when the full 2500-step forecast is considered, the error increases to 2.01, indicating an inaccurate reconstruction of the attractor over the long term. This type of failure is similar to the behavior observed for the ESN forecasts of the fourth system in Figure~\ref{mesncd:fig:comp}b, where the model appeared to be accurate over a short horizon, but eventually deviates from the true trajectory.

        Figures~\ref{mesncd:fig:ts}e and \ref{mesncd:fig:ts}f provide a visual examination of the abrupt failure phenomenon. These panels show the forecast obtained with the CJ topology using a regularization parameter of $\beta = 1.0 \times 10^{-10}$. In the early stages of the prediction, the behavior closely resembles that of the optimal forecast, as shown in the zoomed-in view in Figure~\ref{mesncd:fig:ts}f. The CDE for this initial segment is 0.22. However, around step 800, the forecast shows an abrupt divergence, leading to a catastrophic failure, as can be observed in Figures~\ref{mesncd:fig:ts}e. The final error over the full 900-step time series is 1.78, underscoring the model's failure to reconstruct the chaotic attractor.

\section{\label{mesncd:sec:disc} Discussion and Outlook}

    In this work, we demonstrate that MESNs outperform the randomly initialized ESNs in the task of chaotic attractor reconstruction. Testing on a large dataset of more than 90 chaotic systems, all but one of the minimal topologies yield better reconstruction accuracy than the classic ESN. Furthermore, MESNs exhibit a lower variance in their predictions, resulting in more stable and robust models. We also show that MESNs can share hyperparameters across different systems, enhancing reusability and simplifying model configuration. Finally, we analyze failure modes in ESNs and MESNs, revealing that strong short-term prediction performance does not necessarily imply accurate long-term attractor reconstruction.

    Our findings extend and generalize previous work \citep{ma2023efficient, jaurigue2024chaotic, viehweg2025deterministic} by systematically evaluating multiple minimal deterministic reservoir initializations on a large dataset of chaotic systems. These results show how simpler models can accurately reconstruct chaotic attractors, outshining more complex architectures \citep{schotz2024machine}. Additionally, they emphasize the considerable influence of reservoir topology on attractor reconstruction quality \citep{hemmer2024optimal}. Due to the widespread use of ESNs in disciplines such as Earth sciences \citep{walleshauser2022predicting, hassanibesheli2022longterm, martinuzzi2024learning}, engineering \citep{rackauckas2022composing, anantharaman2022composable, roberts2022continuous}, and time series modeling \citep{kim2020time, bianchi2021reservoir}, exploring simpler and more robust architectural alternatives is necessary to reduce uncertainty and accelerate progress \citep{gauthier2021next}. Our results also indicate the feasibility of constructing simple physical reservoir computing systems \citep{appeltant2011information, tanaka2019recent, nakajima2021scalable, abe2024highly} that can achieve comparable, or even better, performance to their more complex digital counterparts. 

    In this work, we fixed the MESNs reservoir weights at 0.1 for all the experiments. While better values might be found through optimization techniques \citep{fong2025linear}, MESNs still consistently outperformed ESNs, even with a more extensive ESN hyperparameter search. Although the true optima for either model may lie outside the explored search space, this suggests that expanding it would likely not change the relative advantage of MESNs.

    This work opens several directions for future research. One of the main questions is the role of the reservoir weights in minimal deterministic architectures. \citet{fong2025universality} point out that the weight is the sole degree of freedom in such systems, yet we achieved high levels of attractor reconstruction without tuning it. Investigating whether weight optimization yields improved results would be a valuable next step. Another avenue is to explore whether different minimal initializations are better suited to particular classes of chaotic systems. Some topologies may perform consistently well on specific dynamics but fail to do so on others. In this case, “optimization” might shift from tuning hyperparameters to selecting the most appropriate reservoir structure for a given task. This perspective revives a classic question: What makes a good reservoir for a given chaotic system?

\begin{acknowledgments}

    The author thanks Francesca Paoletti, Tristan Williams, and David Montero for helpful discussions regarding the presentation and figures of this work. The author acknowledges the financial support by the Federal Ministry of Research, Technology and Space of Germany and by Sächsische Staatsministerium für Wissenschaft, Kultur und Tourismus in the programme Center of Excellence for AI-research „Center for Scalable Data Analytics and Artificial Intelligence Dresden/Leipzig“, project identification number: ScaDS.AI
    
\end{acknowledgments}

\section*{Author Declaration}

    \subsection*{Conflict of Interest}

        The author has no conflicts to disclose.

    \section*{Author Contributions}

        \textbf{Francesco Martinuzzi}: conceptualization, data curation, formal analysis, methodology, software, validation, visualization, and writing.

\section*{Data Availability}

    The data that support the findings of this study are openly available in \url{https://github.com/MartinuzziFrancesco/mesncd.jl}.



\bibliography{mesncd}

\end{document}